\documentclass[%
 reprint,
 superscriptaddress,
showpacs,
 amsmath,amssymb,
 aps,
 pra,
]{revtex4-1}

\newcommand{\tr}{\textnormal{Tr}}
\newcommand{\ket}[1]{\ensuremath{\left|\right.\!{#1}\!\left.\right\rangle}}
\newcommand{\bra}[1]{\ensuremath{\left\langle\right.\!{#1}\!\left.\right|}}
\newcommand{\braket}[2]{\ensuremath{\langle{#1}|{#2}\rangle}}

\usepackage{amsmath}
\usepackage{graphicx}
\usepackage{dcolumn}
\usepackage{bm}

\begin{document}

\preprint{APS/123-QED}

\title{Quantum walks of interacting fermions on a cycle graph}

\author{Alexey A. Melnikov}
 \email{melnikov@phystech.edu}
 \affiliation{Institute of Physics and Technology, Russian Academy of Sciences, Moscow, Russia}%
\author{Leonid E. Fedichkin}%
 \email{leonid@phystech.edu}
 \affiliation{Moscow Institute of Physics and Technology, Dolgoprudny, Moscow Region, Russia}%

\date{\today}

\begin{abstract}
Quantum walks have been employed widely to develop new tools for quantum information processing recently. A natural quantum walk dynamics of interacting particles can be used to implement efficiently the universal quantum computation. In this work quantum walks of electrons on a graph are studied. The graph is composed of semiconductor quantum dots arranged in a circle. Electrons can tunnel between adjacent dots and interact via Coulomb repulsion, which leads to entanglement. Fermionic entanglement dynamics is obtained and evaluated.
\end{abstract}

\pacs{}

\maketitle


\section{Introduction}

Quantum walks are quantum counterparts of classical random walks~\cite{1993_Aharonov, Aharonov:2001}. Unlike the state of classical walker, quantum walker's state can be a coherent superposition of several positions. Quantum walks found applications to various fields, for example, to the development of a new family of quantum 
algorithms~\cite{shenvi2003quantum, szegedy2004quantum, ambainis2007quantum, KMOR} or to the efficient energy transfer in proteins~\cite{mohseni2008environment}. And recently, quantum walk dynamics is used as an underlying mechanism for quantum-enhanced decision-making process in reinforcement learning~\cite{2012_Briegel_PSI,2014_Paparo_QPS}, for which schemes of experimental realization in systems of trapped ions and superconducting transmon qubits were proposed~\cite{dunjko2015quantum, friis2015coherent}.
There are plenty of theoretical and experimental results in the field of single-particle quantum walks~\cite{venegas2012quantum, zahringer2010realization}, but walks with multiple identical walkers, in both non-interacting and interacting cases, are less explored.

In this paper we study quantum walks of identical particles for quantum information processing purposes. It is known that entanglement creation plays a pivotal role in most of the branches of quantum information. Here we introduce a method for generating a two-qudit (two d-level systems) entangled state by implementing continuous-time quantum walks on a cycle graph. This technique allows us to observe diverse structures of entangled subsystems of high dimensions, a preparation of which is of importance~\cite{Krenn2016, Malik2016}.

To relate theoretical study with feasible experimental implementations we consider realistic models of quantum walks~\cite{wang2014physical}. The physical system we choose as a suitable candidate for quantum walks implementation is an array of tunnel-coupled semiconductor quantum dots. Quantum dots in semiconductors can be used as building blocks for a construction of a quantum computer, where quantum dots positions provide a spatial degree of freedom of a quantum particle~\cite{fedichkin2000coherent, fedichkin2004error, openov2005selective, tsukanov2005entanglement}. It was shown that a spatial location of an electron in one of two semiconductor quantum dots can serve for encoding a qubit~\cite{fedichkin2000coherent,fedichkin2004error} and errors that occur mostly because of the interaction with acoustic phonons can be corrected~\cite{melnikov2013quantum,melnikov2013measure}. In this paper we study quantum dots arranged in a circle, where each quantum dot can be populated by no more than one electron. By placing two identical particles in this system, one can define higher dimensional quantum states, qudits. If electrons are close enough they can also influence each other via Coulomb interaction.

The remainder of the paper has the following structure: first, we introduce the model of symmetrical two-electron quantum walk on a cycle graph of arbitrary size. After the introduction of the model we study the dynamics of electrons in the cases with and without an interaction between them. The case of interacting electrons is studied in details and the scheme for entangling gate between two qudits, represented by two electrons, is proposed.
Then we summarize the results and discuss possible applications of the proposed scheme.

\section{Framework}

The system under consideration contains two electrons. Each electron can sit in one of $N$ quantum dots arranged in a circle~\cite{solenov2006continuous}. Dots themselves can be formed from the two-dimensional electron gas by field of gates and the population of electrons in these dots can be controlled by potentials on gates. Each position in the circle can be occupied by at most one electron. The position of an electron can be measured by quantum point contact detectors, which are placed near quantum dots such that an electron in a certain quantum dot decreases an electric current in the detector by increasing a potential barrier. Therefore a lower current detects an electron and a higher current indicates an absence of an electron (an empty quantum dot), correspondingly.

Experimentally, lateral structures of this geometry with different number of quantum dots were realized. Among them, a double quantum dot, which can be viewed as a circle with $N=2$ sites, is the most studied configuration and is used to create a solid state qubit~\cite{RevModPhys.75.1,RevModPhys.79.1217,NanoLettCoupling,RevModPhys.85.961}. Beyond this, triple quantum dots with circular and linear geometries were studied in detail both theoretically and experimentally~\cite{RepProgPhys3dot}. A concept of a scalable architecture was demonstrated by fabricating quadruple~\cite{APL4dots,APL4dots2,APL4dots3} and quintuple~\cite{ito2016detection} quantum dots. In all experiments, a high degree of control over the precise number of electrons in each quantum dot was demonstrated by measuring stability diagrams. Moreover, it was shown that it is possible to tune the tunnel coupling between neighbouring quantum dots by changing the voltage on the gate that spatially separates these dots, see e.g. Ref.~\cite{NanoLettCoupling}, where the tunnel coupling was shown to be an exponential function of the gate voltage. Similar techniques and technologies could be used for a fabrication of circles of larger sizes.

\begin{figure}[h]
\center{\includegraphics[width=.55\linewidth]{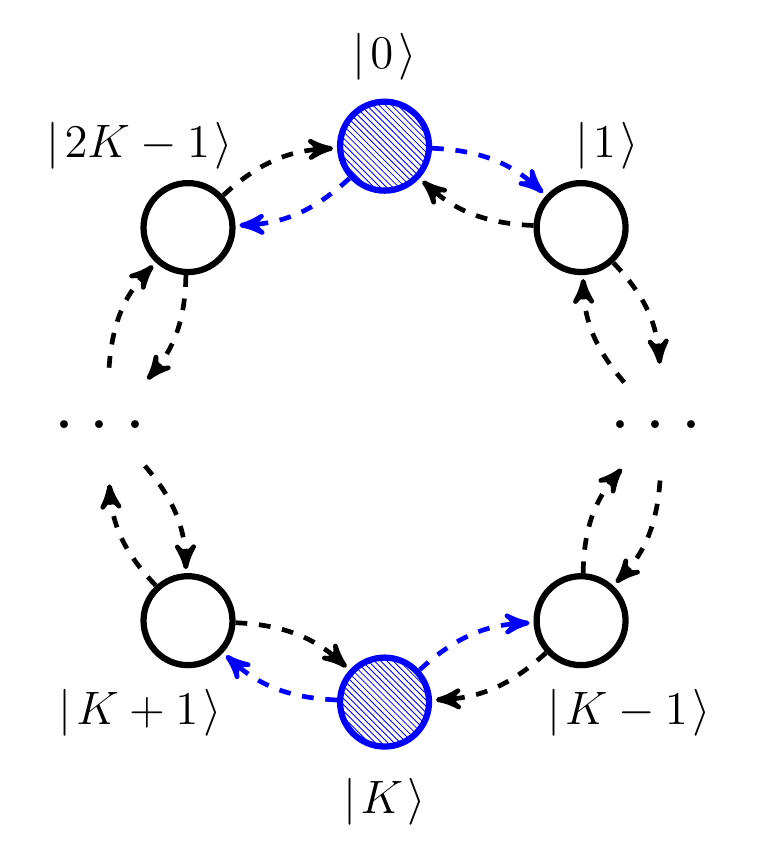}}
\vspace{-3mm}\caption{A cycle graph with $N = 2K$ vertices, where each vertex is viewed as a level in the $N$-level quantum system. Two electrons are initially placed in the $ 0 $-th and $ K $-th positions, i.e. to an initial $\ket{\Psi(0)} = \left(\ket{K0}-\ket{0K}\right)\ket{\uparrow\uparrow}/\sqrt{2}$ state. This initial state is chosen in order to achieve a high symmetry in this system.}
\label{fig:Nnodes}
\end{figure}

The described circle of semiconductor quantum dots is mathematically represented as a cycle graph with quantum dots being vertices of this graph. Edges of the cycle graph connect only nearest neighbours and represent possible tunnel transitions of electrons. We enumerate the vertices within the graph, from $0$ to $N-1$. The localization of an electron in the $ 0 $-th, $ 1 $-st, $\dots $ or $(N-1) $-th quantum dot is described by corresponding quantum states $ \ket 0 $, $ \ket 1$, $\dots$ or $ \ket{N-1}$, as shown in Fig.~\ref{fig:Nnodes} for $N=2K$, $K\in\mathbb{N}$. As a straightforward result, the states $ \ket 0 $, $ \ket 1$, ... and $ \ket{N-1}$ can be viewed as the basis states of a qudit, whose amplitudes squared correspond to the probabilities of detecting an electron. Note that because electrons cannot occupy the same energy level, i.e. the same vertex on a cycle graph, $\ket{ii}$ two-qudit basis states are impossible for all $i \in[0,~2K-1]$.

\vspace{-3mm}

Electrons are initially placed in opposed vertices of the graph, as depicted in Fig.~\ref{fig:Nnodes}, but can later change their positions by hopping between neighbouring vertices. This process is a continuous-time quantum walk governed by the Hamiltonian, which we introduce below. Electrons walk and spread due to tunneling through the barrier of controlled height between the quantum dots. 
For the sake of simplicity we assume that electrons spins are always up ($ 1/2 $), which can be the case, for example, in a strong magnetic field. Wave function of two indistinguishable fermions in form of $ \ket{\Psi (t)} = \ket{\psi(t)} \ket{\uparrow\uparrow} $, an antisymmetric coordinate part of which is
\begin{align} 
\ket{\psi(t)} & = \sum_{m, k=0}^{N-1}\omega_{mk}(t) \ket{m, k} \nonumber\\
 & = \frac{1}{\sqrt{2}}\sum_{m, k=0}^{N-1}\omega_{mk}(t) \ket{\psi^{(m, k)}},
\label{eq:psi2}
\end{align}
where $\ket{m, k}$ is the state with the first and second electron being in the $m$-th and $k$-th vertex, respectively; $\ket{\psi^{(m, k)}}$ is the state of electrons occupying vertices $m$ and $k$, which corresponds to a product state of two uncorrelated systems. These electrons have to be treated as indistinguishable because their wave functions overlap spatially in the quantum dots. The state in Eq.~(\ref{eq:psi2}) is a superposition of electrons being in different vertices with time-dependent amplitudes $ \omega_{mk}(t) $, which form a matrix $\omega (t)$~\cite{eckert2002quantum}. The matrix $\omega (t)$ is antisymmetric, i.e. $\omega^T (t) = -\omega (t)$, and takes into account the antisymmetric nature of the fermionic wave function. The normalization of the $\ket{\psi(t)}$ state gives an additional condition on $\omega (t)$:
\begin{equation}
\mathrm{Tr}\left( \omega (t) \omega^\dag (t)\right)  = 1.
\end{equation}

The wave function specified by $\omega$ matrix fully characterizes a fermionic state, however a correct definition of its subsystems is required for studying properties of the system. The problem of reduced fermionic density operators was addressed recently in Refs.~\cite{friis2015reasonable, amosov2015spectral}, where it is shown that parity superselection rule should be applied to the fermionic state and the unique definition of the reduced density operator is provided. In our case, the total number of fermions is constant, which leads to the standard procedure of obtaining the reduced state $\rho_1 (t)$:
\begin{align} 
\rho_1 (t) & = \mathrm{Tr}_2 \Big(\ket{\psi(t)}\bra{\psi(t)}\Big) \nonumber\\
 & = \mathrm{Tr}_2 \Bigg(\sum_{m, m', k, k'=0}^{N-1}\omega_{mk}(t) \ket{m, k}\bra{m', k'} \omega^*_{m'k'}(t)\Bigg) \nonumber\\
 & = \sum_{m, m', k, k'=0}^{N-1}\omega_{mk}(t)\omega^*_{m'k'}(t)~~\mathrm{Tr}_2 \Big( \ket{m, k}\bra{m', k'} \Big) \nonumber\\ 
 & = \sum_{m, m', k=0}^{N-1}\omega_{mk}(t)\omega^*_{m'k}(t) \ket{m}\bra{m'}.
\label{eq:pTrace}
\end{align}
This definition is used later to study the entanglement properties of the system.

\section{Non-interacting indistinguishable electrons}

The dynamics of the two-electron fermionic state depends on an arrangement of quantum dots: if quantum dots are close enough -- electrons will interact through Coulomb repulsion, otherwise electrons do not interact. First, we consider the case without interaction and move to the case with interaction afterwards.

The evolution of an electron in an array of tunnel-coupled semiconductor quantum dots can be modelled by a continuous-time quantum walk, which is defined by a Hamiltonian with nearest-neighbour interactions~\cite{solenov2006continuous}. By analogy, the evolution of two electrons can be modelled by a continuous-time quantum walk of two particles that is governed by the Hamiltonian
\begin{align} 
H_\mathrm{O} & = \hbar\Omega \sum_{m, k = 0}^{N-1} \Big( \ket{(m+1)~\mathrm{mod}~N,~k}\bra{m,~k} + \nonumber\\
 & + \ket{k,~(m+1)~\mathrm{mod}~N}\bra{k,~m} + \mathrm{H.c.} \Big),
\label{eq:HamilNonint}
\end{align}
where $ \Omega $ is the tunneling frequency, which corresponds to the potential barrier height between neighbouring quantum dots. This Hamiltonian is defined for $N>2~(K>1)$, for the smallest graph ($K=1$) the Hamiltonian is equal to the half of the one in Eq.~\ref{eq:HamilNonint}, i.e. $H_\mathrm{O}/2$, since in the circle of two dots clockwise and counterclockwise jumps correspond to the same transition. One can see, that the Hamiltonian $H_\mathrm{O}$ only changes the spatial part of the total fermionic wave function $\ket{\Psi (t)}$, leaving the spin part unchanged. In other words, the walk is performed in the space of coordinates of quantum dots, and the spins remain parallel as they were initially prepared. Therefore, the spin part of the wave function factors out from the evolution and will not be taken into account below. The remaining part of the total wave function, the antisymmetric spatial part, evolves according to the Schr\"{o}dinger equation $ \ket{\psi(t)} = e^{-iH_\mathrm{O}t/\hbar} \ket{\psi(0)} $, where the unitary operator $e^{-iH_\mathrm{O}t/\hbar}$ can be shown to map any antisymmetric fermionic wave function to antisymmetric one. An exact matrix representation of this unitary operator can be obtained analytically for small $K$, but in general can only be computed numerically. 

In Methods we provide exact solutions of the Schr\"{o}dinger equation for $K=2, 3, 4$. Exact solutions let us observe the periodic dynamics for $K = 2$ and $3$ with periods $T = \pi/2\Omega$ and $2\pi/3\Omega$, respectively, and aperiodic dynamics for $K= 4$ (see Methods for details). From these results we conclude that in general the dynamics is aperiodic, as it was also shown in the case of discrete-time quantum walks on cycles~\cite{Dukes2014189, konno2015periodicity}. Although the dynamics is aperiodic, it is known that by waiting enough time, an arbitrary precision of returning to the initial state can be achieved, as shown in Methods for $K=4$. The possibility to achieve an arbitrary precision of the state revival holds for all $K$ and is known from the Poincar\'e recurrence theorem~\cite{bocchieri1957quantum, wallace2015recurrence}, although in general for different $K$ it might take different time to achieve the same level of precision.

\begin{figure*}[t!]
\center{
\includegraphics[width=1\linewidth]{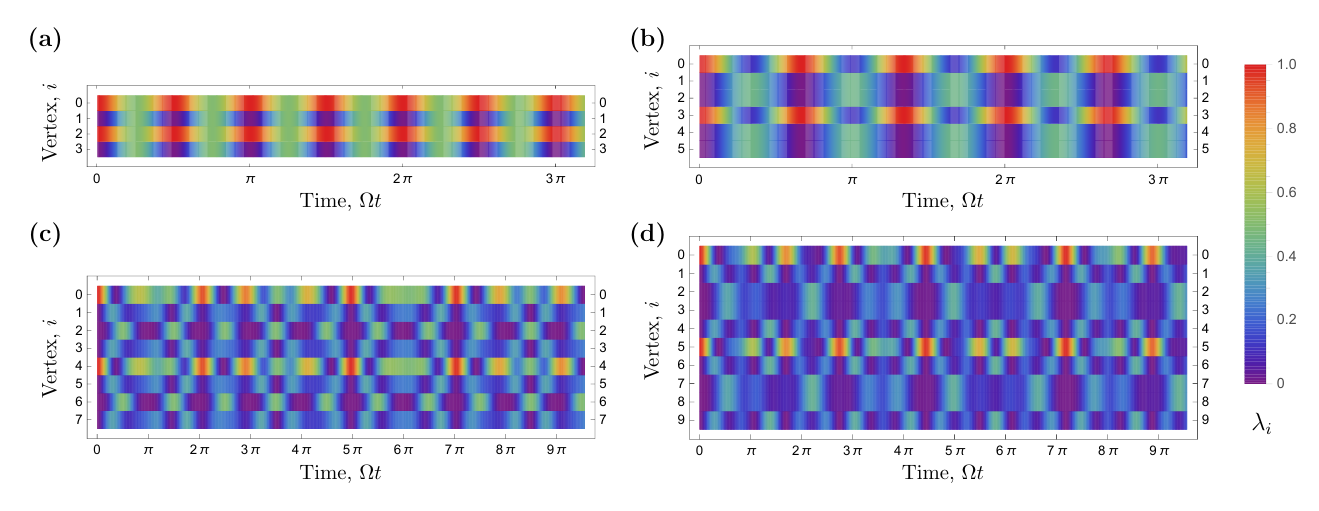}
}
\vspace{-2mm}\caption{The average number of electrons $\lambda_i$ in a vertex vs. time. The initial state is $ \ket{\Psi (0)} = ( \ket{K0} - \ket{0K} )\ket{\uparrow\uparrow}/\sqrt{2} $. (a) Quantum walk dynamics for $K=2$, initial state is fully recovered after the time $ \pi/2\Omega $. (b) Quantum walk dynamics for $K=3$, initial state is fully recovered after the time $ 2\pi/3\Omega $. (c) Quantum walk dynamics for $K=4$, initial state is partially recovered after the time $\Omega t = 3\pi/\sqrt{2} \approx 2.12\pi$, $7\pi/\sqrt{2} \approx 4.95\pi$ and $10\pi/\sqrt{2} \approx 7.07\pi$. (d) Quantum walk dynamics for $K=5$, initial state is partially recovered after the time $\Omega t\approx 2.8\pi$, $4.4\pi$ and $7.2\pi$.}
\label{fig:dynamics4nodes}
\end{figure*}

In experiment, the wave function $\ket{\psi(t)}$ cannot be directly observed, the measured data corresponds to a population in each quantum dot, i.e an average number of electrons in each dot. For this reason our function of interest is the population $\lambda_i$ in the vertex $i$ of the cycle graph. The population $\lambda_i$ is equal to the probability to detect an electron in the vertex $i$ and is related to the amplitudes $\omega_{mk}$ of the wave function $\ket{\psi(t)}$:
\begin{equation}
\lambda_i = \sum_{k=0}^{N-1} \omega_{ki}\omega_{ki}^* + \omega_{ik}\omega_{ik}^* = 2\sum_{k=0}^{N-1} |\omega_{ik}|^2.
\label{eq:averageElectrons}
\end{equation}
Fig.~\ref{fig:dynamics4nodes} shows population dynamics $\lambda_i(t)$ for the smallest $K=2, 3, 4$ and $5$. The solution in the case of $K=1$ is trivial $ \ket{\psi(t)} = \ket{\psi(0)} $ and is not shown. From Fig.~\ref{fig:dynamics4nodes}(a) and (b) one can immediately deduce that the charge dynamics is periodic, confirming the analytical results for the periods of quantum walks $T=\pi/2\Omega$ and $2\pi/3\Omega$ in the case of $K=2$ (a) and $K=3$ (b), respectively. The dynamics in the case of $K=4$ (c) is, however, aperiodic, which is proven in Methods. But as we discussed before, a nearly full state revival can be observed in this system, in particular in the case of $K=4$ (c) and $K=5$ (d).

A population distribution dynamics $\lambda'_i(t) = \lambda_i(t)/2$, similar to the one shown in Fig.~\ref{fig:dynamics4nodes}, can be obtained by having only one electron initially prepared in a superposition of $ \ket{0} $ and $ \ket{K} $ coordinate states, where the scaling factor of $1/2$ comes from the reduction of the total charge in the system. This can be seen from the right part of Eq.~\ref{eq:averageElectrons} -- the position of the second particle $k$ is irrelevant, the distribution $\lambda_i$ only depends on the position of the first particle. The probability to find this single electron in a certain node is half of the probability of finding one of two non-interacting electrons, which is also a consequence of Eq.~\ref{eq:averageElectrons}. Hence a quantum walk of two non-interacting particles can be simulated by a one-particle walk, whose dynamics was studied in Refs.~\cite{solenov2006continuous, fedichkin2006mixing, solenov2006nonunitary}. But because it is not straightforward to initialize an electron in a superposition of being in different nodes, two-particle walk can be used for studying one-particle walks with arbitrary initial conditions.

\section{Interacting indistinguishable electrons}

Here we consider the case of two identical electrons that interact through Coulomb repulsion. The mutual repulsion between electrons becomes apparent when the distance between the quantum dots is such that the emerged Coulomb energy induced by one of the electrons prevents the second electron to tunnel to the adjacent dot. In order to model a fermionic quantum walk we approximate the Coulomb interaction by restricting the positions of electrons: electrons cannot be in the same or neighbouring vertices of the graph, and the effect of repulsion is negligible in all other situations, i.e. an electron does not ``feel'' an electric field of the distant electrons, if the distance between them is more than one empty quantum dot. 
This approximation is reasonable because neigbour dots are generally closer to each other than to metallic gates forming them so interaction can be strong, while interaction of electrons at distant dots is substantially suppressed not only by larger distance of interaction but also by screening due to presence of metallic gates between and nearby them.
The Hamiltonian with the restriction of not being in the same and neighbouring vertices of the cycle graph is
\begin{widetext}
\begin{equation} 
H_\mathrm{C} = \hbar\Omega \sum_{k = 0}^{N-1} \sum_{m = k+2}^{N+k-3} \Big( \ket{(m+1)~\mathrm{mod}~N,~k}\bra{m~\mathrm{mod}~N,~k} + \ket{k,~(m+1)~\mathrm{mod}~N}\bra{k,~m~\mathrm{mod}~N} + \mathrm{H.c.} \Big).
\label{eq:HamilInt}
\end{equation}
\end{widetext}

\begin{figure*}[t!]
\center{\includegraphics[width=1\linewidth]{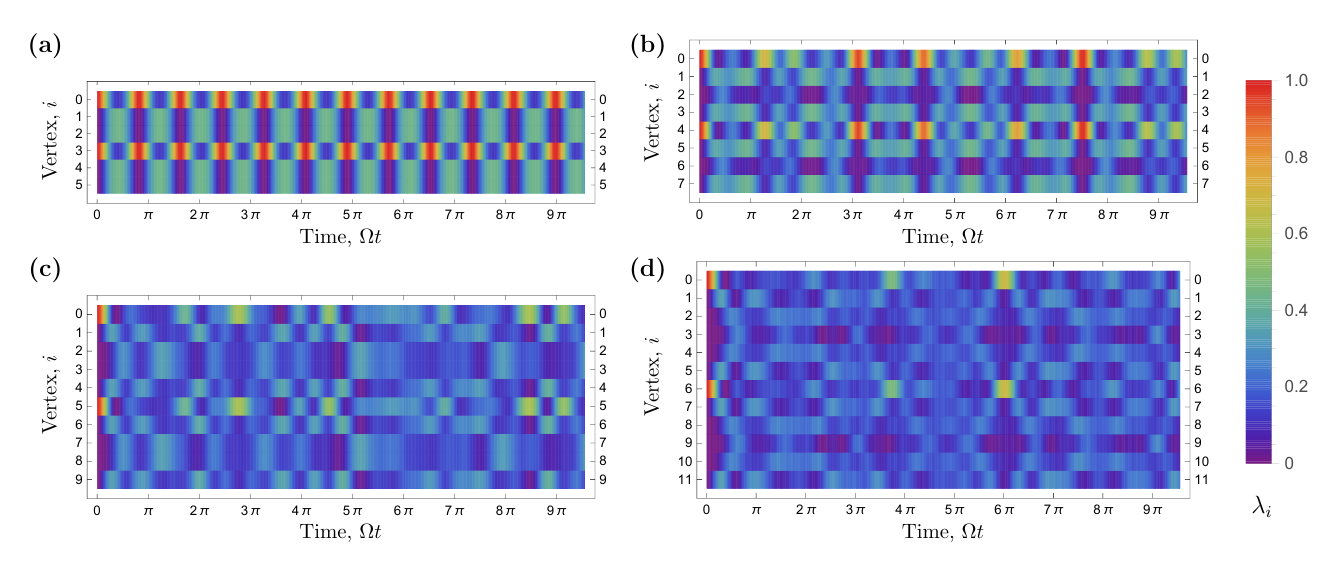}}
\vspace{-2mm}\caption{The average number of electrons $\lambda_i$ in a vertex vs. time. Mutual repulsion between electrons is taken into account. The initial state is $ \ket{\Psi (0)} = ( \ket{K0} - \ket{0K} )\ket{\uparrow\uparrow}/\sqrt{2} $. (a) Quantum walk dynamics for $K=3$, initial state is fully recovered after the time $ 2\pi/\sqrt{6}\Omega $. (b) Quantum walk dynamics for $K=4$, initial state is partially recovered after the time $\Omega t = 10\pi/\sqrt{6+3\sqrt{2}} \approx 3.1\pi$ and $\Omega t = 24\pi/\sqrt{6+3\sqrt{2}} \approx 7.5\pi$. (c) Quantum walk dynamics for $K=5$, initial state is partially recovered after the time $\Omega t \approx 4.6\pi$, $8.5\pi$. (d) Quantum walk dynamics for $K=6$, initial state is partially recovered after the time $\Omega t \approx 3.9\pi$, $6.0\pi$.}
\label{fig:dynamics8nodes}
\end{figure*}

Similar to the case of non-interacting electrons, we obtain analytical solutions of the Schr\"{o}dinger equation $ \ket{\psi(t)} = e^{-iH_\mathrm{C}t/\hbar} \ket{\psi(0)} $ for small dimensions of the cycle graph $K=2$, $3$ and $4$ (the case of $K=1$ is unfeasible, because it is impossible to place two strongly repelling electrons in two quantum dots). The results, provided in Methods, demonstrate that there exists a period of quantum walks for $K=3$, but not for $K=4$. Hence, in general, the quantum walk of interacting particle on a cycle aperiodic. This fact can be seen in the population dynamics $\lambda_i(t)$ plotted in Fig.~\ref{fig:dynamics8nodes} for $K=3$, $4$, $5$ and $6$.

\subsection{Fermionic entanglement by means of a quantum walk}

It is known that interactions between particles create quantum entanglement between these particles~\cite{PhysRevLett.86.910,PhysRevLett.82.1975}. The qualification and quantification of an entanglement between several subsystems is one of the most important issues in quantum information theory. However, by describing an entanglement of two fermions we cannot use the standard definition of entanglement of distinguishable particles, because for identical particles the Hilbert space has no longer a tensor product structure. More specifically, the Hilbert space of two electrons is an antisymmetric product, not a direct product~\cite{gittings2002describing, zanardi2002quantum}.

To define entanglement of indistinguishable fermions one can use the Slater rank~\cite{schliemann2001quantum, eckert2002quantum, chernyavskiy2009entanglement}. The Slater rank is the minimum number of Slater determinants, and this number is an analogue of the Schmidt rank for the distinguishable case. Fermions are called separable iff the Slater rank is equal to one. That is quantum entanglement arise in a pure state if there is no single-particle basis such that a given state of electrons can be represented as a single Slater determinant
\begin{equation}
\ket{\psi^{(m, k)}} = \frac{1}{\sqrt{2}} \big( \ket{m}\otimes\ket{k} - \ket{k}\otimes\ket{m} \big).
\end{equation}

Fermionic quantum correlations defined above are the analogue of quantum entanglement between distinguishable systems and are essential for quantum information processing with indistinguishable systems. However these correlations should be quantified differently from the case of distinguishable systems by taking into account the definition of fermionic entanglement. Defining good measures of fermionic entanglement remains a field of active research~\cite{PhysRevA.92.042326,PhysRevA.93.032335}. In this paper we use three fermionic entanglement measures: von Neumann entropy~\cite{plastino2009separability,zander2012entropic}, linear entropy~\cite{plastino2009separability,zander2012entropic} and fermionic concurrence~\cite{PhysRevA.93.032335}.

Von Neumann entropy of the pure state $\rho=\ket{\psi}\bra{\psi}$ is
\begin{equation}
S_\mathrm{vN}(\rho) = -\mathrm{Tr}(\rho_1\ln \rho_1) - \ln 2 = -\sum_{j} \xi_j\ln\xi_j - \ln 2,
\label{eq:SvN}
\end{equation}
where $\rho_1$ is the single-particle reduced density matrix defined in Eq.~\ref{eq:pTrace}, and $\xi_j$ are the nonzero eigenvalues of the $\rho_1$ matrix. It was shown, that a pure state $\rho$ has the Slater rank equal to one iff $ S_\mathrm{vN}(\rho) = 0 $~\cite{eckert2002quantum, amico2008entanglement, buscemi2007linear}. An entanglement criterion for states of two fermions can also be formulated in terms of the linear entropy
\begin{equation}
S_\mathrm{L}(\rho) = \frac{1}{2} - \mathrm{Tr}\rho_1^2,
\label{eq:SL}
\end{equation}
which is the approximation of the von Neumann entropy. A pure state $\rho$ has the Slater rank equal to one iff $ S_\mathrm{L}(\rho) = 0 $. 
We also use the fermionic concurrence~\cite{PhysRevA.93.032335} 
\begin{equation}
C_\mathrm{f}(\rho) = \sqrt{\frac{2N}{N-2}\bigg( \frac{1}{2}-\tr{\rho^2_1}\bigg)},
\label{eq:Cf}
\end{equation}
which by analogy with the linear entropy gives $0$ for separable states and nonzero values for entangled fermionic states. In addition, the fermionic concurrence in Eq.~\ref{eq:Cf} is normalized between $0$ and $1$.

\begin{figure*}[t!]
\center{\includegraphics[width=.9\linewidth]{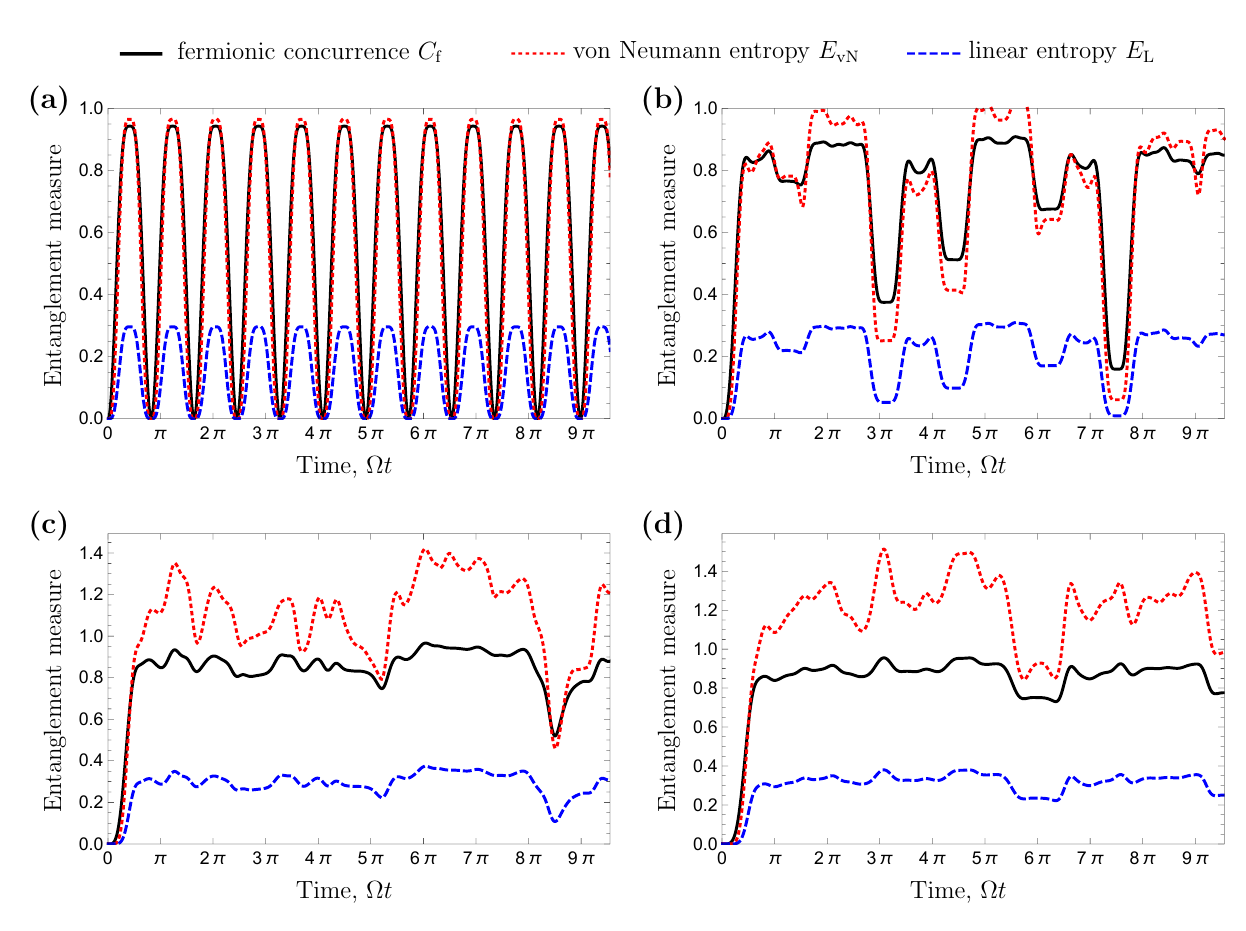}}
\caption{Entanglement measures: fermionic concurrence $C_\mathrm{f}$ (solid black), von Neumann (dotted red) and linear (dashed blue) entropy. (a) $K=3$. Entanglement exhibits periodic dynamics with the period $T=2\pi/\sqrt{6}\Omega$. (b) $K=4$. Entanglement dynamics is aperiodic. At time $\Omega t \approx 7\pi$ there is a sudden drop of entanglement with the local minimum of entanglement at time $ \Omega t \approx 7.5\pi $. (c) $K=5$. At time $ \Omega t \approx 8.5\pi$ there is a drop of entanglement because of the partial revival of the initial state. (d) $K=6$. At time $ \Omega t \approx 6.0\pi$ there is a drop of entanglement because of the partial revival of the initial state.}
\label{fig:entanglement}
\end{figure*}

We calculate $ S_\mathrm{vN}(\rho(t)) $, $ S_\mathrm{L}(\rho(t)) $ and $C_\mathrm{f}(\rho(t))$ functions using Eqs.~\ref{eq:SvN},~\ref{eq:SL} and~\ref{eq:Cf}, respectively, for $K=3, 4, 5$ and $6$. These entanglement measures are shown in Fig.~\ref{fig:entanglement}. It can be seen that two electrons are initially separable, but after a time, which increases with $K$, they become entangled. In the case of $K=3$, shown in Fig~\ref{fig:entanglement}(a), the entanglement dynamics is periodic, as expected due to the periodicity of the wave function. The maximum entanglement is achieved at times $t=\pi (1+2n)/\sqrt{6}\Omega$, $n\in\mathbb{N}$, for the state
\begin{equation}
\ket{\psi_3} = -\frac{1}{3}\left(2\ket{\psi^{(1, 4)}} + 2\ket{\psi^{(5, 2)}} + \ket{\psi^{(0, 3)}} \right).
\label{eq:K3ent}
\end{equation}
The minimum entanglement corresponds to the separable state $\ket{\psi^{(0, 3)}}$, which is present at times $t = 2\pi n/\sqrt{6}\Omega$, $n\in\mathbb{N}$.

The evolution of entanglement for $K=4$ ($N=8$ vertices) is shown in Fig~\ref{fig:entanglement}(b). One can see that the particles entangle slower (initial slope in Fig.~\ref{fig:entanglement}) than in case of $K=3$, because electrons are initially further away from each other and it takes more time for particles to meet each other. The evolution is aperiodic and the entanglement never disappears, but because of a partial revival of the initial separable state, the entanglement of electrons drops suddenly at times of the largest overlap with the initial state $\Omega t \approx 3.1\pi$ and $\Omega t \approx 7.5\pi$ (see also Fig.~\ref{fig:dynamics8nodes}). The maximum entanglement is achieved for multiple states. For instance, at times $\Omega t = 7\pi/\sqrt{6+3\sqrt{2}} \approx 2.2\pi$, $\Omega t = 17\pi/\sqrt{6+3\sqrt{2}} \approx 5.3\pi$ and $\Omega t = 27\pi/\sqrt{6+3\sqrt{2}} \approx 8.4\pi$ the following fermionic state is generated
\begin{equation}
\ket{\psi_4} = -\frac{1}{3}\left(2\ket{\psi^{(1, 3)}} + 2\ket{\psi^{(7, 5)}} - \ket{\psi^{(0, 4)}} \right).
\label{eq:K4ent}
\end{equation}

Fig.~\ref{fig:entanglement}(c) and (d) show the entanglement dynamics for higher cycle graph dimensions $K=5$ and $6$, respectively. Similar to the case of $K=4$, the dynamics is aperiodic and maximal entanglement is achieved for many states. There are also occasional drops of entanglement caused by a partial return to the initial state $\ket{\psi^{(0, K)}}$.

The fermionic entanglement initiation described here is due to the Coulomb interaction. This repulsive interaction can be interpreted as a condition that restricts the positions of electrons -- a quantum walk of one electron is conditioned on the state of the second electron and vice versa. In contrast to the interacting case, non-interacting electrons do not have this conditioned dynamics; dynamics of electrons is independent from each other. It can easily be shown that in absence of this Coulomb repulsion condition, entanglement is not initiated and all mentioned fermionic entanglement measures are equal to zero throughout the entire quantum walk evolution. Indeed, the Hamiltonian in Eq.~\ref{eq:HamilNonint} that governs the evolution of non-interacting electrons, leads to an independent unitary dynamics of two electrons
\begin{widetext}
\begin{align} 
\ket{\psi(t)} & = \mathrm{e}^{-i\Omega t \sum_{m, k = 0}^{N-1} \big( \ket{(m+1)~\mathrm{mod}~N}\bra{m} + \mathrm{H.c.} \big)\otimes \ket{k}\bra{k}} \mathrm{e}^{-i\Omega t \sum_{m, k = 0}^{N-1} \ket{m}\bra{m} \otimes \big( \ket{(k+1)~\mathrm{mod}~N}\bra{k} + \mathrm{H.c.} \big)}\ket{\psi(0)} \nonumber\\
 & = \left(\mathrm{e}^{-i\Omega t \sum_{m = 0}^{N-1} \big( \ket{(m+1)~\mathrm{mod}~N}\bra{m} + \mathrm{H.c.} \big)}\otimes I\right)\left( I \otimes\mathrm{e}^{-i\Omega t \sum_{k = 0}^{N-1} \big( \ket{(k+1)~\mathrm{mod}~N}\bra{k} + \mathrm{H.c.} \big)}\right)\ket{\psi(0)} \nonumber\\
 & = U_\mathrm{O}(t)\otimes U_\mathrm{O}(t)\ket{\psi^{(0, K)}},
\end{align}
\end{widetext}
where by $U_\mathrm{O}(t)$ we denote a time-dependent unitary matrix that acts locally on a subspace of one particle. Because the initial state of two electrons is separable, local operations clearly cannot create an entangled state. To verify this we compute the reduced density state from Eq.~\ref{eq:pTrace}:
\begin{widetext}
\begin{align} 
\rho_1 (t) & = \frac{1}{2}\Big(U_\mathrm{O}(t)\ket{0}\bra{0}U^{\dag}_\mathrm{O}(t) - U_\mathrm{O}(t)\ket{K}\bra{0}U^{\dag}_\mathrm{O}(t)\bra{K} U_\mathrm{O}(t)U^{\dag}_\mathrm{O}(t)\ket{0} - U_\mathrm{O}(t)\ket{0}\bra{K}U^{\dag}_\mathrm{O}(t)\bra{0} U_\mathrm{O}(t)U^{\dag}_\mathrm{O}(t)\ket{K}\nonumber\\
 & + U_\mathrm{O}(t)\ket{K}\bra{K}U^{\dag}_\mathrm{O}(t)\Big) = \frac{1}{2}~U_\mathrm{O}(t)\big(\ket{0}\bra{0} + \ket{K}\bra{K}\big)U^{\dag}_\mathrm{O}(t).
\end{align}
\end{widetext}
Using the explicit form of the reduced density matrix $\rho_1 (t)$ that we obtained, we compute $\mathrm{Tr}\rho_1^2(t) = 1/2$ and $\mathrm{Tr}(\rho_1(t)\ln \rho_1(t)) = -\ln 2$, which leads us to the observation that all entanglement measures, $S_\mathrm{vN}(\rho, t)$, $S_\mathrm{L}(\rho, t)$ and $C_\mathrm{f}(\rho, t)$ are equal to zero for all times $t$. Because these measures are zero iff the fermionic state $\rho$ is separable, we conclude that, as expected, only by allowing electrons to interact, one can introduce fermionic entanglement in the system of coupled quantum dots that we consider.

\subsection{Electrons for quantum information processing}

We showed that a variety of entangled fermionic states can be created by means of quantum walks. However, it is not apparent how useful are these fermionic states for quantum information processing, because of inconsistency between fermions and qubits~\cite{friis2015reasonable}. Here we show how one can define qudits by using the freedom of dividing the graph into two subgraphs. We divide the cycle graph into two equal parts: the first subgraph contains the vertices $\{0, \dots, \lfloor K/2\rfloor, \lfloor K/2\rfloor+ K +1, \dots, 2K-1\}$, the second subgraph contains the vertices $\{\lfloor K/2\rfloor +1, \dots, \lfloor K/2\rfloor+K\}$. Experimentally, this division can be realized by raising a potential barrier between the two pairs of quantum dots, $\lfloor K/2\rfloor$ and $\lfloor K/2\rfloor +1$ dots, and between $\lfloor K/2\rfloor+K$ and $\lfloor K/2\rfloor+ K +1$ dots. As we demonstrate below, in our framework, due to the symmetry of the initial state, it is possible to see that electrons are confined in different subgraphs with the unit probability. In this case, we say that an electron in the upper subgraph (vertices $0, \dots, \lfloor K/2\rfloor, \lfloor K/2\rfloor+ K +1, \dots, 2K-1$) and an electron in the lower subgraph (vertices $\lfloor K/2\rfloor +1, \dots, \lfloor K/2\rfloor+K$) represent two distinguishable qudits. Below we show that this definition of two qudits in terms of the upper and the lower subspaces allows obtaining highly entangled states of two qudits.

\begin{figure*}[t!]
\center{\includegraphics[width=0.95\linewidth]{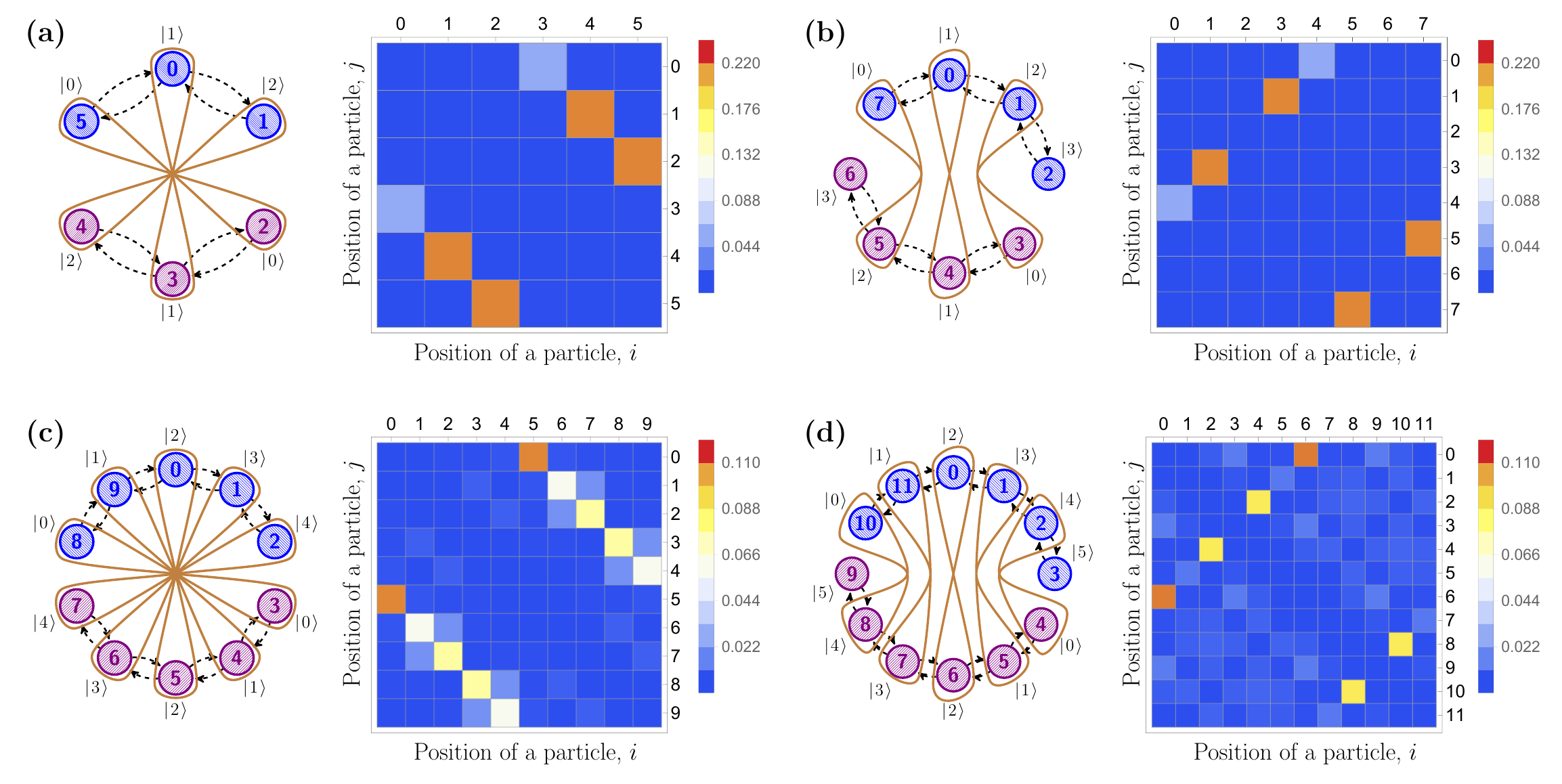}}
\caption{The left part of each figure (a) -- (d) shows a scheme of the cycle graph with $2K$ vertices divided into two subgraphs, each of which represents a state space of a qudit. The subspace of the first qudit is shown in blue (vertices $0$, $\dots$, $\lfloor K/2\rfloor$, $\lfloor K/2\rfloor+ K +1$, $\dots$, $2K-1$), the subspace of the second qubit is shown in violet (vertices $\lfloor K/2\rfloor +1$, $\dots$, $\lfloor K/2\rfloor+K$). Dashed black arrows show the possible transitions between the vertices in the redefined graph. Brown curves represent the type of entanglement (see text for details). The right part of each figure (a) -- (d) shows a matrix of two-particle correlations of quantum walkers in position space. The element ($j$,$i$) of the correlations matrix corresponds to a probability of detecting two electrons in quantum dots $j\in\{0,\dots,N-1\}$ and $i\in\{0,\dots,N-1\}$. (a) $K=3$. The entangled two-qutrit state corresponds to a fermionic state obtained at times $t=\pi (1+2n)/\sqrt{6}\Omega$, $n\in\mathbb{N}$. (b) $K=4$. The entangled two-ququart state corresponds to a fermionic state obtained at time $t = 17\pi/\sqrt{6+3\sqrt{2}}\Omega $. (c) $K=5$. The entangled two-qudit state corresponds to a fermionic state obtained at time $t=24.3/\Omega$. (d) $K=6$. The entangled two-qudit state corresponds to a fermionic state obtained at time $t=25.7/\Omega$.}
\label{fig:CircleCut}
\end{figure*}

The described ``cuts" of the circle are depicted in Fig.~\ref{fig:CircleCut} (a) -- (d) for $K=3$ (a), $K=4$ (b), $K=5$ (c) and $K=6$ (d). Fig.~\ref{fig:CircleCut}(a) schematically shows two qutrits \big(three-level systems with basis states $\ket{0}$, $\ket{1}$ and $\ket{2}$\big) defined on a cycle graph. At time $t=\pi/\sqrt{6}\Omega$, as we have shown before, the quantum dynamics on a cycle graph with $6$ vertices leads to the state in Eq.~\ref{eq:K3ent} with two-particle correlation matrix shown in the right part of Fig.~\ref{fig:CircleCut} (a). One can see, that if one raises a potential barrier between quantum dots $1$ and $2$, $4$ and $5$, as shown in the left part of Fig.~\ref{fig:CircleCut} (a), one traps electrons in separate subgraphs, because at this time the particles can only be detected in the stictly opposite sites at the circle. After defining qudits at this time step we obtain an entangled state
\begin{equation}
\ket{\psi_3} = -\frac{1}{3}\left(2\ket{00} + \ket{11} + 2\ket{22} \right).
\label{eq:K3entNew}
\end{equation}
The brown curves in Fig.~\ref{fig:CircleCut}~(a) represent the type of a superposition in Eq.~\ref{eq:K3entNew}, which corresponds to the Bell-type entanglement.

Fig.~\ref{fig:CircleCut} (b) depicts a larger cycle graph with $6$ vertices. Similar to the case of $K=3$, two qudits are defined on the circle at a certain time $t = 17\pi/\sqrt{6+3\sqrt{2}}\Omega$. At this time one can separate two halves of the circle and obtain the following state of two ququarts
\begin{equation}
\ket{\psi_4} = -\frac{1}{3}\left(2\ket{02} + 2\ket{20} - \ket{11} \right).
\label{eq:K4entNew}
\end{equation}
Although this state is similar to the one in Eq.~\ref{eq:K3entNew} up to a local phase, the type of entanglement in space of the graph is different and shown in the left part of Fig.~\ref{fig:CircleCut}~(b). From the right part of the Fig.~\ref{fig:CircleCut}~(b) one can see that electrons are distributed in different subgraphs. The cases of $K=5$ and $K=6$ are shown in Fig.~\ref{fig:CircleCut} (c) and (d), respectively. As one can see from correlation matrices, the same types of quantum correlations can be achieved with high probability.

The fermionic entanglement dynamics studied in this paper is obtained for a pure state of a quantum system. However, in experiment the quantum state is subjected to decoherence. In particular, a change in the state of the qudits can be caused by the white noise from the quantum point contact detectors, which was shown to be one of the major concerns facing experimental realization of quantum walks in quantum dots structures~\cite{solenov2006continuous}. This noise can be modelled by a depolarizing channel that acts on a density matrix of two fermions $\rho$ as follows
\begin{equation}
\rho(t) = e^{-\Gamma t}\rho(0)+\left(1-e^{-\Gamma t}\right)\rho_M,
\label{eq:decoherence}
\end{equation}
which is the solution of the differential equation $ d\rho(t)/dt = -\Gamma(\rho(t)-\rho_M) $ with the initial state $\rho(0) = \ket{\psi^{(0, K)}}\bra{\psi^{(0, K)}}$, where $ \Gamma $ is the relaxation rate that corresponds to the coupling between the quantum dot and the quantum point contact. The density matrix $\rho_M$ is the maximally mixed state of the coordinate part of two electrons. Because of the antisymmetric fermionic state the maximally mixed state is not the normalized identity matrix, but is defined as follows
\begin{widetext}
\begin{equation}
\rho_M = \dfrac{1}{N(N-3)}\sum_{k=0}^{N-1}~\sum_{m=k+2}^{N+k-2} \ket{\psi^{(m~\mathrm{mod}~N,~k)}} \bra{\psi^{(m~\mathrm{mod}~N,~k)}}.
\end{equation}
\end{widetext}
Combining the dissipative dynamics from Eq.~\ref{eq:decoherence} with the coherent evolution with Hamiltonian $H_\mathrm{C}$ from Eq.~\ref{eq:HamilInt} we write the general expression for the fermionic density matrix
\begin{align} 
\rho(t) & = e^{-iH_\mathrm{C} t/\hbar} \left(e^{-\Gamma t}\rho(0)+\left(1-e^{-\Gamma t}\right)\rho_M \right) e^{iH_\mathrm{C} t/\hbar} \nonumber\\
 & = e^{-\Gamma t}e^{-iH_\mathrm{C}t/\hbar}\rho(0)e^{iH_\mathrm{C}t/\hbar}+\left(1-e^{-\Gamma t}\right)\rho_M.
\end{align}
We are able to analyze the combination of the two different processes and to do the simplification of the expression due to the relation $\left[H_\mathrm{C}, \rho_M\right] = 0$, which implies that the operator $e^{-iH_\mathrm{C}t/\hbar}$ commutes with $\rho_M$. 

\begin{figure}
\center{\includegraphics[width=1\linewidth]{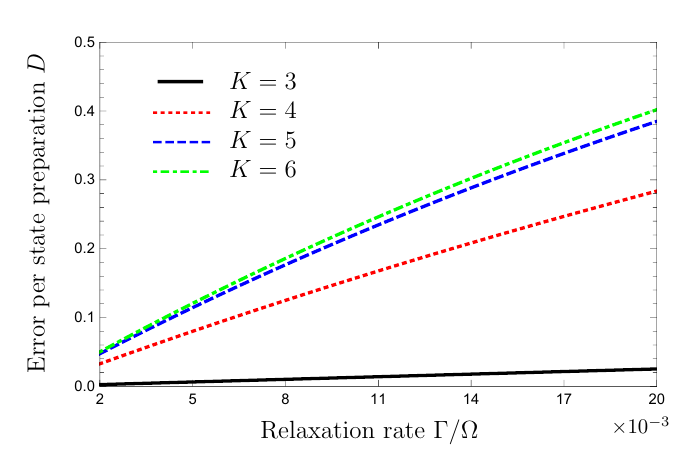}}
\caption{The dependance of an error per state preparation on the relaxation rate is depicted. Four curves correspond to the states shown in Fig.~\ref{fig:CircleCut} (a) -- (d).}
\label{fig:decoherence}
\end{figure}

The decoherence described by Eq.~(\ref{eq:decoherence}) leads to errors in quantum information stored in the system of electrons. In order to quantify this error we use the measure of decoherence~\cite{fedichkin2004additivity, melnikov2013measure}
\begin{align} 
D & = \left|\left|\rho_{\rm{real}} - \rho_{\rm{ideal}}\right|\right| \nonumber\\
 & = \left(1 - e^{-\Gamma t}\right)\left|\left|e^{-iH_\mathrm{C}t/\hbar}\rho(0)e^{iH_\mathrm{C}t/\hbar} - \rho_M\right|\right|
\end{align}
to quantify the amount of errors, where the operator norm of the matrix $ X $ is given by $ \left|\left|X\right|\right|=\max_{\substack{x\in \mathrm{spec}(X)}}\left|x\right| $ with $ \mathrm{spec}(X) $ being the spectrum of the operator $ X $. The measure of decoherence $D$ can be thought of as a probability of obtaining an error. In our scenario getting an error would correspond to getting a completely classical state of two particles, which are uniformly distributed over the circle, instead of getting entangled states shown in Fig.~\ref{fig:CircleCut}. As shown in Fig.~\ref{fig:decoherence}, where we plot an error for each state in the set from Fig.~\ref{fig:CircleCut} (entangled states with $K=3$, $4$, $5$ and $6$), this error can become large for large circles and strong couplings $\Gamma$ between the quantum dot and the quantum point contact.

In addition to the described noise, the system of electrons is subjected to the inevitable phase noise caused by the deformation interaction of electrons with acoustic phonons~\cite{fedichkin2004error}. As a result, the energy levels in quantum dots where electrons reside become not fully determined, which effectively distorts the nondiagonal elements of the density matrix $\widetilde{\rho}$ as follows
\begin{equation}
\widetilde{\rho} = \sum_{i=0}^K E_i\otimes I\left(\sum_{j=0}^K I\otimes E_j\ket{\psi_K}\bra{\psi_K} I\otimes E_j\right) E_i\otimes I,
\label{eq:dephasing}
\end{equation}
where $I$ is the identity matrix, $E_i = \sqrt{1-e^{-\gamma}}\ket{i}\bra{i}$ for $i<K$ and $E_K = e^{-\gamma/2}I$ with $\gamma = \Xi^2/2\hbar\pi^2 \rho s^3 a^2$. Note that the $\ket{\psi_K}$ state is written in the basis of separated electrons, which corresponds to the states in Eqs.~\ref{eq:K3entNew} and~\ref{eq:K4entNew}. The following parameters are taken for electrons in silicon: effective deformation potential $\Xi = 3.3$ eV, speed of sound $s = 9.0\times 10^3$ m/s, density $\rho = 2.33$~g/cm$^3$, and quantum dot size $a = 10$ nm~\cite{fedichkin2004error}. For this set of parameters and $K=4$ the obtained phase error is $D \approx 1.4\times 10^{-5}$, which suggests that this additional phase error is negligible and the error is mostly determined by the depolarizing noise in the range of relaxation rates we consider in Fig.~\ref{fig:decoherence}.

\section{Conclusion}
In the paper we considered the dynamics of two-particle fermionic system. We analyzed quantum walks in two possible setups, which lead to a walk with and without interaction between electrons. We preferred quantum walks approach to quantum information processing for a number of reasons. 
Quantum walks dynamics is a natural process for many quantum systems compared to more artificial gate implementation. It is therefore easier to build and implement in experiment. One may also hope that relatively complex gates sequences could be replaced by simpler quantum walks processes.  It was shown before that one can do arbitrary quantum operations using only particles free propagation~\cite{venegas2012quantum}. One way to realize quantum walks algorithms is to use silicon quantum dots that form a cycle graph. We showed the electrons entanglement dynamics in this structure. The value of fermionic entanglement was calculated using measures in Eq.~(\ref{eq:SvN}), (\ref{eq:SL}) and (\ref{eq:Cf}), which were proven to correctly quantify entanglement~\cite{plastino2009separability,zander2012entropic,PhysRevA.93.032335}. We showed that fermionic entanglement can be used to prepare quantum states for quantum information processing. These highly entangled states of qudits can be obtained by only using the free quantum evolution of identical particles, without relying on any additional manipulations with electrons. In addition, we supplemented our protocol of obtaining entangled states with analytical solutions for certain sizes of a graph and proved a general aperiodic nature of the continuous-time quantum walk of identical particles on a cycle graph.

\begin{acknowledgments}
We wish to thank Sergey N. Filippov for helpful discussions. The work of L.~E.~F. is supported by Russian Science Foundation under grant No.~16-01-00084 and performed in Moscow Institute of Physics and Technology.
\end{acknowledgments}

\section*{Methods}
\label{sec:methods}

In this section we explain the fermionic quantum walks dynamics in details by obtaining explicit analytical solutions of the Schr\"{o}dinger equation. Throughout the section we use the series expansion of the quantum walk unitary operator
\begin{equation}
\ket{\psi(t)} = \ket{\psi(0)} + \sum_{l=1}^{\infty} \frac{(-it/\hbar)^l}{l!} H^l\ket{\psi(0)},
\label{eq:PeriodSum}
\end{equation}
where $H$ is $H_\mathrm{O}$ in case of non-interacting electrons and $H_\mathrm{C}$ in case of interacting electrons. This expansion is useful in the case of a continuous-time quantum walk on a circle, because due to the cyclic conditions the number of fermionic states that can be observed is bounded.

\subsection*{Period of quantum walks of non-interacting particles}

We first start our analysis with the case of quantum walks of non-interacting indistinguishable electrons, whose dynamics is described by the Hamiltonian $H_\mathrm{O}$ in Eq.~\ref{eq:HamilNonint}. We first consider the smallest sizes of the cycle graph with $K=1$ ($2$ vertices), $K=2$ ($4$ vertices), $K=3$ ($6$ vertices) and show the periodicity of the underlying dynamics. Next we show that, in general, the dynamics is aperiodic, i.e. there is no time $T\neq0$ s.t. $\ket{\psi(T)}=\ket{\psi(0)}$, by obtaining the solution for $K=4$ ($8$ vertices).

\textit{A cycle graph with 2 vertices}.
For $K=1$ the evolution of the state is trivial: $ \ket{\psi(t)} = e^{-iH_\mathrm{O}t/\hbar} \ket{\psi(0)} = \ket{\psi(0)} $, because the initial state is the eigenstate of the Hamiltonian $H_\mathrm{O}$. This is expected due to the fact that the system of two quantum dots has only two energy levels both occupied by electrons, and because of the Pauli exclusion principle these electrons cannot change their positions.

\textit{A cycle graph with 4 vertices}.
By computing the lower powers of $H_\mathrm{O}$ for $K=2$ we observe that $H_\mathrm{O}^2 \ket{\psi^{(0,2)}} = 4\hbar^2\Omega^2 \ket{\psi^{(0,2)}}$. Hence, using the observation we reduce Eq.~\ref{eq:PeriodSum} for this size of the graph to
\begin{widetext}
\begin{align} 
\ket{\psi(t)} & = \sum_{l=0}^\infty \frac{(-it/\hbar)^{2l}}{(2l)!} (2\hbar\Omega)^{2l}\ket{\psi^{(0,2)}} + \sum_{l=0}^\infty \frac{(-it/\hbar)^{2l+1}}{(2l+1)!} (2\hbar\Omega)^{2l}H_\mathrm{O}\ket{\psi^{(0,2)}} \nonumber\\
 & = \cos{(2\Omega t)}\ket{\psi^{(0,2)}} - \frac{i}{2\hbar\Omega}\sin{(2\Omega t)}H_\mathrm{O}\ket{\psi^{(0,2)}} \nonumber\\
 & = \cos{(2\Omega t)}\ket{\psi^{(0,2)}} - \frac{i}{2}\sin{(2\Omega t)}\left[\ket{\psi^{(0,1)}} + \ket{\psi^{(0,3)}} + \ket{\psi^{(1,2)}} + \ket{\psi^{(3,2)}}\right].
\end{align}
\end{widetext}
An overlap of the state $\ket{\psi(t)}$ with the initial state is equal to $\left|\braket{\psi(0)}{\psi(t)}\right| = \left|\cos{(2\Omega t)}\right|$, therefore the dynamics of the wave function (as well as the population $\lambda_i$ and the fermionic entanglement functions) is periodic. The period of the dynamics is $T = \pi/2\Omega$, after this time the initial state is fully revived (we neglect a global phase). It is worth noting, that at time $t=\pi/\Omega$ the unitary matrix $e^{-iH_\mathrm{O} t/\hbar}$ is equal to the identity matrix $I$, so any initial state is recovered after this time. The specific choice of the symmetric initial state we use recovers twice more frequent.

\textit{A cycle graph with 6 vertices}.
Similarly to the case of $K=2$, we compute the lower powers of $H_\mathrm{O}$ and obtain the relation $H_\mathrm{O}^3 \ket{\psi_3(0)} = 9\hbar^2\Omega^2 H_\mathrm{O}\ket{\psi_3(0)}$, which leads us to the state
\begin{widetext}
\begin{align} 
\ket{\psi(t)} & = \ket{\psi(0)} + \sum_{l=1}^\infty \frac{(-it/\hbar)^{2l}}{(2l)!} (3\hbar\Omega)^{2l-2}H_\mathrm{O}^2\ket{\psi(0)} + \sum_{l=0}^\infty \frac{(-it/\hbar)^{2l+1}}{(2l+1)!} (3\hbar\Omega)^{2l}H_\mathrm{O}\ket{\psi(0)} \nonumber\\
 & = \ket{\psi(0)} + \frac{1}{9\hbar^2\Omega^2}(\cos{(3\Omega t)}-1)H_\mathrm{O}^2\ket{\psi(0)} - \frac{i}{3\hbar\Omega}\sin{(3\Omega t)}H_\mathrm{O}\ket{\psi(0)} = \ket{\psi^{(0,3)}} + \frac{1}{9}(\cos{(3\Omega t)}-1) \nonumber\\
 & \phantom{= } \times\left[ \ket{\psi^{(0,1)}} + \ket{\psi^{(0,5)}} + \ket{\psi^{(2,3)}} + \ket{\psi^{(4,3)}} + 4\ket{\psi^{(0,3)}} + 2\left( \ket{\psi^{(1,2)}} + \ket{\psi^{(1,4)}} + \ket{\psi^{(5,2)}} + \ket{\psi^{(5,4)}}\right)\right] \nonumber\\
 & \phantom{= } - \frac{i}{3}\sin{(3\Omega t)}\left[ \ket{\psi^{(0,2)}} + \ket{\psi^{(0,4)}} + \ket{\psi^{(1,3)}} + \ket{\psi^{(5,3)}} \right].
\end{align}
\end{widetext}
An overlap of the state $\ket{\psi(t)}$ with the initial state is equal to $\left|\braket{\psi(0)}{\psi(t)}\right| = \left|5 + 4\cos{(3\Omega t)}\right|/9$, therefore the dynamics is periodic with the period $T = 2\pi/3\Omega$. It is worth noting, that at time $t=2\pi/\Omega$ unitary matrix $e^{-iH_\mathrm{O} t/\hbar}$ is equal to identity matrix $I$, so any initial state is recovered after this time. The specific choice of the symmetric initial state we use recovers $3$ times more frequent.

\textit{A cycle graph with 8 vertices}.
The case of $K=4$ is already more involved. We first divide the sum from Eq.~\ref{eq:PeriodSum} in two sums with even and odd powers of $H_\mathrm{O}$, respectively:
\begin{align} 
\ket{\psi(t)} & = \sum_{l=0}^\infty \frac{(-it/\hbar)^{2l}}{(2l)!} H_\mathrm{O}^{2l}\ket{\psi^{(0,4)}} \nonumber\\
 & + \sum_{l=0}^\infty \frac{(-it/\hbar)^{2l+1}}{(2l+1)!} H_\mathrm{O}^{2l+1}\ket{\psi^{(0,4)}},
\label{eq:Period4non-int}
\end{align}
where
\begin{align} 
& \left(H_\mathrm{O}/\hbar\Omega\right)^{2l}\ket{\psi^{(0,4)}} = \alpha_1^{(l)} \Big( \ket{\psi^{(0,2)}} + \ket{\psi^{(0,6)}} \nonumber\\
 & + \ket{\psi^{(2,4)}} + \ket{\psi^{(6,4)}}\Big) + \alpha_2^{(l)} \Big( \ket{\psi^{(1,3)}} + \ket{\psi^{(1,5)}} \nonumber\\
 & + \ket{\psi^{(7,3)}} + \ket{\psi^{(7,5)}}\Big) + \alpha_3^{(l)} \ket{\psi^{(0,4)}}
 \label{eq:Period41non-int}
\end{align}
and

\begin{align} 
& \left(H_\mathrm{O}/\hbar\Omega\right)^{2l+1}\ket{\psi^{(0,4)}} = \beta_1^{(l)} \Big( \ket{\psi^{(0,1)}} + \ket{\psi^{(0,7)}} + \ket{\psi^{(3,4)}} \nonumber\\
& + \ket{\psi^{(5,4)}}\Big) + \beta_2^{(l)} \Big( \ket{\psi^{(1,2)}} + \ket{\psi^{(1,6)}} + \ket{\psi^{(2,3)}} \nonumber\\
& + \ket{\psi^{(2,5)}} + \ket{\psi^{(6,3)}} + \ket{\psi^{(6,5)}} + \ket{\psi^{(7,2)}} + \ket{\psi^{(7,6)}}\Big) \nonumber\\
& + \beta_3^{(l)} \Big( \ket{\psi^{(0,3)}} + \ket{\psi^{(0,5)}} + \ket{\psi^{(1,4)}} + \ket{\psi^{(7,4)}}\Big).
 \label{eq:Period42non-int}
\end{align}
An application of the $H_\mathrm{O}^{2l}$ to the unnormalized states in Eqs.~\ref{eq:Period41non-int} and~\ref{eq:Period42non-int} preserves their structure, and only changes the coefficients $\alpha_i^{(l)}$ and $\beta_i^{(l)}$, respectively:

\begin{align}
\left(
\begin{array}{c}
\alpha_1^{(l)}\\
\alpha_2^{(l)}\\
\alpha_3^{(l)}\\
\end{array}
\right)
& =
\left(
\begin{array}{ccc}
4 & 4 & 1\\
4 & 6 & 2\\
4 & 8 & 4\\
\end{array}
\right)^l
\left(
\begin{array}{c}
0\\
0\\
1\\
\end{array}
\right), \nonumber\\
\left(
\begin{array}{c}
\beta_1^{(l)}\\
\beta_2^{(l)}\\
\beta_3^{(l)}\\
\end{array}
\right)
& =
\left(
\begin{array}{ccc}
1 & 2 & 1\\
1 & 4 & 3\\
1 & 6 & 9\\
\end{array}
\right)^l
\left(
\begin{array}{c}
0\\
0\\
1\\
\end{array}
\right).
\label{eq:Period4non-intCoef}
\end{align}
By using the relations from Eqs.~\ref{eq:Period41non-int}, \ref{eq:Period42non-int} and~\ref{eq:Period4non-intCoef} we compute the sum in Eq.~\ref{eq:Period4non-int} and obtain the solution of the Schr\"{o}dinger equation
\begin{widetext}
\begin{align} 
& \ket{\psi(t)} = -\frac{1}{4}\cos{\left(\sqrt{2}\Omega t\right)}\left(1-\cos{\left(2\Omega t\right)}\right) \left( \ket{\psi^{(0,2)}} + \ket{\psi^{(0,6)}} + \ket{\psi^{(2,4)}} + \ket{\psi^{(6,4)}}\right) \nonumber\\
&- \frac{1}{2\sqrt{2}}\sin{\left(\sqrt{2}\Omega t\right)}\sin{\left(2\Omega t\right)} \left( \ket{\psi^{(1,3)}} + \ket{\psi^{(1,5)}} + \ket{\psi^{(7,3)}} + \ket{\psi^{(7,5)}}\right) + \frac{1}{2}\cos{\left(\sqrt{2}\Omega t\right)}\left(1+\cos{\left(2\Omega t\right)}\right) \ket{\psi^{(0,4)}} \nonumber\\
&- \frac{i}{4\sqrt{2}}\left(\sqrt{2}\cos{\left(\sqrt{2}\Omega t\right)}\sin{\left(2\Omega t\right)} - \sin{\left(\sqrt{2}\Omega t\right)}(1+\cos{\left(2\Omega t\right)}) \right) \left( \ket{\psi^{(0,1)}} + \ket{\psi^{(0,7)}} + \ket{\psi^{(3,4)}} + \ket{\psi^{(5,4)}}\right) \nonumber\\
&+ \frac{i}{4\sqrt{2}}\sin{\left(\sqrt{2}\Omega t\right)}\left(1-\cos{\left(2\Omega t\right)}\right) \Big( \ket{\psi^{(1,2)}} + \ket{\psi^{(1,6)}} + \ket{\psi^{(2,3)}} + \ket{\psi^{(2,5)}} + \ket{\psi^{(6,3)}} + \ket{\psi^{(6,5)}} + \ket{\psi^{(7,2)}} + \ket{\psi^{(7,6)}}\Big) \nonumber\\
&- \frac{i}{4\sqrt{2}}\left(\sqrt{2}\cos{\left(\sqrt{2}\Omega t\right)}\sin{\left(2\Omega t\right)} + \sin{\left(\sqrt{2}\Omega t\right)\left(1+\cos{\left(2\Omega t\right)}\right)} \right) \left( \ket{\psi^{(0,3)}} + \ket{\psi^{(0,5)}} + \ket{\psi^{(1,4)}} + \ket{\psi^{(7,4)}}\right).
\end{align}
\end{widetext}
An overlap of the state $\ket{\psi(t)}$ with the initial state is $\left|\braket{\psi(0)}{\psi(t)}\right| = \frac{1}{2}\left|\cos{\left(\sqrt{2}\Omega t\right)}\right|\left(1+\cos{\left(2\Omega t\right)}\right)$. This overlap is unit only when $t=0$, therefore there is no period of quantum walks for $K=4$. However, by choosing the time $t$ s.t. $\cos{\left(\sqrt{2}\Omega t \right)} \approx \pm 1 $ and $ \cos{\left(\Omega t \right)} \approx \pm 1$, i.e. $\sqrt{2}\Omega t = \pi n$, $n \approx \sqrt{2}k$ with $n, k\in \mathbb{Z}$, the overlap $\left|\braket{\psi(0)}{\psi(t)}\right|$ is close to unity. By waiting enough, an arbitrary precision can be achieved.

\subsection*{Period of quantum walks of interacting particles}
We next move to the case of interacting particles, which quantum dynamics is described by Eq.~\ref{eq:PeriodSum} with the Hamiltonian $H_\mathrm{C}$. The minimal graph size in the case of repulsive electrons is $K=2$, for which the dynamics is trivial with stationary solution $ \ket{\psi(t)} = e^{-iH_\mathrm{C}t/\hbar} \ket{\psi(0)} = \ket{\psi(0)} $.

\textit{A cycle graph with 6 vertices}.
By computing the lower powers of the $H_\mathrm{C}$ for $K=3$, we see that  $H_\mathrm{C}^3 \ket{\psi(0)} = 6\hbar^2\Omega^2 H_\mathrm{C}\ket{\psi(0)}$, hence 
\begin{widetext}
\begin{align} 
& \ket{\psi(t)} = \ket{\psi(0)} + \sum_{l=1}^\infty \frac{(-it/\hbar)^{2l}}{(2l)!} (\sqrt{6}\hbar\Omega)^{2l-2}H_\mathrm{C}^2\ket{\psi(0)} + \sum_{l=0}^\infty \frac{(-it/\hbar)^{2l+1}}{(2l+1)!} (\sqrt{6}\hbar\Omega)^{2l}H_\mathrm{C}\ket{\psi(0)} \nonumber\\
& = \ket{\psi(0)} + \frac{1}{6\hbar^2\Omega^2}(\cos{(\sqrt{6}\Omega t)}-1)H_\mathrm{C}^2\ket{\psi(0)} - \frac{i}{\sqrt{6}\hbar\Omega}\sin{(\sqrt{6}\Omega t)}H_\mathrm{C}\ket{\psi(0)} = \ket{\psi^{(0,3)}} + \frac{1}{3}(\cos{(\sqrt{6}\Omega t)}-1) \nonumber\\
& \times\left[ \ket{\psi^{(1,4)}} + \ket{\psi^{(5,2)}} + 2 \ket{\psi^{(0,3)}}\right] - \frac{i}{\sqrt{6}}\sin{(\sqrt{6}\Omega t)}\left[ \ket{\psi^{(0,2)}} + \ket{\psi^{(0,4)}} + \ket{\psi^{(1,3)}} + \ket{\psi^{(5,3)}} \right].
\end{align}
\end{widetext}
An overlap of the obtained solution with the initial state is $\left|\braket{\psi(0)}{\psi(t)}\right| = \left|\frac{1}{3} + \frac{2}{3}\cos{(\sqrt{6}\Omega t)}\right|$. Therefore the dynamics of the wave function is periodic with the period $T = 2\pi/\sqrt{6}\Omega$, which means that after the time $T$ the initial state is fully revived.

\textit{A cycle graph with 8 vertices}. Similar to the case of non-interacting electrons, the dynamics for the size $K=4$ is more involved. We first decompose the sum from Eq.~\ref{eq:PeriodSum} in the following way:
\begin{align} 
\ket{\psi(t)} & = \sum_{l=0}^\infty \frac{(-it/\hbar)^{2l}}{(2l)!} H_\mathrm{C}^{2l}\ket{\psi^{(0,4)}} \nonumber\\
 & + \sum_{l=0}^\infty \frac{(-it/\hbar)^{2l+1}}{(2l+1)!} H_\mathrm{C}^{2l+1}\ket{\psi^{(0,4)}},
\label{eq:Period4int}
\end{align}
where
\begin{align} 
& \left(H_\mathrm{C}/\hbar\Omega\right)^{2l}\ket{\psi^{(0,4)}} = \alpha_1^{(l)} \Big( \ket{\psi^{(0,2)}} + \ket{\psi^{(0,6)}} \nonumber\\
&  + \ket{\psi^{(2,4)}} + \ket{\psi^{(6,4)}}\Big) + \alpha_2^{(l)} \left( \ket{\psi^{(1,3)}} + \ket{\psi^{(7,5)}}\right) \nonumber\\
& + \alpha_3^{(l)} \left( \ket{\psi^{(1,5)}} + \ket{\psi^{(7,3)}} \right) + \alpha_4^{(l)} \ket{\psi^{(0,4)}}
\label{eq:Period41}
\end{align}
and
\begin{align} 
& \left(H_\mathrm{C}/\hbar\Omega\right)^{2l+1}\ket{\psi^{(0,4)}} = \beta_1^{(l)} \Big( \ket{\psi^{(1,6)}} + \ket{\psi^{(2,5)}} \nonumber\\
& + \ket{\psi^{(6,3)}} + \ket{\psi^{(7,2)}} \Big) + \beta_2^{(l)} \Big( \ket{\psi^{(0,3)}} + \ket{\psi^{(0,5)}} \nonumber\\
& + \ket{\psi^{(1,4)}} + \ket{\psi^{(7,4)}}\Big).
\label{eq:Period42}
\end{align}
An application of the $H_\mathrm{C}^{2l}$ to the unnormalized states in Eq.~\ref{eq:Period41} and~\ref{eq:Period42} preserves their structure, and only changes the coefficients $\alpha_i^{(l)}$ and $\beta_i^{(l)}$, respectively:

\begin{align}
\left(
\begin{array}{c}
\alpha_1^{(l)}\\
\alpha_2^{(l)}\\
\alpha_3^{(l)}\\
\alpha_4^{(l)}\\
\end{array}
\right)
& =
\left(
\begin{array}{cccc}
2 & 1 & 2 & 1\\
2 & 2 & 2 & 2\\
4 & 2 & 4 & 2\\
4 & 4 & 4 & 4\\
\end{array}
\right)^l
\left(
\begin{array}{c}
0\\
0\\
0\\
1\\
\end{array}
\right), \\\nonumber
\left(
\begin{array}{c}
\beta_1^{(l)}\\
\beta_2^{(l)}\\
\end{array}
\right)
& =
\left(
\begin{array}{ccc}
3 & 3\\
3 & 9\\
\end{array}
\right)^l
\left(
\begin{array}{c}
0\\
1\\
\end{array}
\right).
\label{eq:Period4intCoef}
\end{align}
By using the relations from Eqs.~\ref{eq:Period41}, \ref{eq:Period42} and~\ref{eq:Period4intCoef} we compute the sum in Eq.~\ref{eq:Period4int} and obtain the solution of the Schr\"{o}dinger equation
\begin{widetext}
\begin{align} 
& \ket{\psi(t)} = \frac{\cos{(\omega_+ t)}-\cos{(\omega_- t)}}{6\sqrt{2}} \left( \ket{\psi^{(0,2)}} + \ket{\psi^{(0,6)}} + \ket{\psi^{(2,4)}} + \ket{\psi^{(6,4)}} + 2\ket{\psi^{(1,5)}} + 2\ket{\psi^{(7,3)}} \right) \nonumber\\
& + \frac{\cos{(\omega_+ t)}+\cos{(\omega_- t)}}{6} \left( \ket{\psi^{(1,3)}} + \ket{\psi^{(7,5)}} + 2\ket{\psi^{(0,4)}}\right) + \frac{1}{3} \left( \ket{\psi^{(0,4)}} -\ket{\psi^{(1,3)}} - \ket{\psi^{(7,5)}}\right) \nonumber\\
& + i~\frac{\omega_+\sin(\omega_-t) - \omega_-\sin(\omega_+t)}{12\Omega} \Big( \ket{\psi^{(1,6)}} + \ket{\psi^{(2,5)}} + \ket{\psi^{(6,3)}} + \ket{\psi^{(7,2)}} \Big) \nonumber\\
& - i~\frac{\omega_+\sin(\omega_+t) + \omega_-\sin(\omega_-t)}{12\Omega} \left( \ket{\psi^{(0,3)}} + \ket{\psi^{(0,5)}} + \ket{\psi^{(1,4)}} + \ket{\psi^{(7,4)}}\right),
\end{align}
\end{widetext}
where $\omega_+ = \sqrt{6+3\sqrt{2}}~\Omega$ and $\omega_- = \sqrt{6-3\sqrt{2}}~\Omega$. 
An overlap of the obtained solution with the initial state is $\left|\braket{\psi(0)}{\psi(t)}\right| = \frac{1}{3}\left|1 + \cos{(\omega_+ t)} + \cos{(\omega_- t)} \right|$. This overlap is unit only when $t=0$, therefore there is no period of quantum walks for $K=4$. However an arbitrary precision of state revival can be achieved by choosing the time $t$ s.t. $\cos{(\omega_+ t)}\approx \cos{(\omega_- t)} \approx 1$, i.e. $t = 2\pi n/\omega_+$, $k \approx n\omega_-/\omega_+ = (\sqrt{2}-1)n$ with $n, k\in \mathbb{Z}$. This happens approximately, e.g. for $\Omega t = 10\pi\Omega/\omega_+ \approx 3.1\pi$ ($n=5$, $k\approx 2$) and $\Omega t = 24\pi\Omega/\omega_+ \approx 7.5\pi$ ($n=12$, $k\approx 5$).

\providecommand{\noopsort}[1]{}\providecommand{\singleletter}[1]{#1}%

\end{document}